\documentclass[namedreferences]{SolarPhysics}
\usepackage{epsfig}

\usepackage{graphicx}

\newcommand{\be}{\begin{equation}}
\newcommand{\ee}{\end{equation}}
\newcommand{\beq}{\begin{eqnarray}}
\newcommand{\eeq}{\end{eqnarray}}

\newcommand{\arcsec}{^{\prime \prime}}

\begin{document}
\begin{article}
\begin{opening}

\title{%
DO QUASI-REGULAR STRUCTURES REALLY EXIST IN THE SOLAR PHOTOSPHERE?
\\I. Observational evidence}

\author{A.V. GETLING}

\runningauthor{A.V. GETLING}

\runningtitle{QUASI-REGULAR STRUCTURES IN THE PHOTOSPHERE I.}

\institute{%
Institute of Nuclear Physics, Lomonosov Moscow State University,
119992 Moscow, Russia
\\ \email{A.Getling@ru.net} }
\date{Received May 29, 2006; accepted *** 00, 2006}

\begin{abstract}
Two series of solar-granulation images --- the La Palma series of
5 June 1993 and the SOHO MDI series of 17--18 January 1997 --- are
analysed both qualitatively and quantitatively. New evidence is
presented for the existence of long-lived, quasi-regular
structures (first reported by Getling and Brandt (2002)), which no
longer appear unusual in images averaged over 1--2-h time
intervals. Such structures appear as families of light and dark
concentric rings or families of light and dark parallel strips
(``ridges'' and ``trenches'' in the brightness distributions). In
some cases, rings are combined with radial ``spokes'' and can thus
form ``web'' patterns. The characteristic width of a ridge or
trench is somewhat larger than the typical size of granules.
Running-average movies constructed from the series of images are
used to seek such structures. An algorithm is developed to obtain,
for automatically selected centres, the radial distributions of
the azimuthally averaged intensity, which highlight the
concentric-ring patterns. We also present a time-averaged
granulation image processed with a software package intended for
the detection of geological structures in aerospace images. A
technique of running-average-based correlations between the
brightness variations at various points of the granular field is
developed and indications are found for a dynamical link between
the emergence and sinking of hot and cool parcels of the solar
plasma. In particular, such a correlation analysis confirms our
suggestion that granules --- overheated blobs --- may repeatedly
emerge on the solar surface. Based on our study, the critical
remarks by Rast \cite{rast} on the original paper by Getling and
Brandt (2002) can be dismissed.
\end{abstract}
\end{opening}

\section{Introduction} \label{Introduction}
As reported previously by Getling \& Brandt (2002; hereinafter,
paper I), the procedure of time averaging applied to a 2-h
interval of the 11-h La Palma series of granulation images (see
below) reveals signs of long-lived, quasi-regular photospheric
structures --- ``ridges'' and ``trenches'' in the brightness
distributions, which form systems of concentric rings or parallel
strips. These systems resemble some roll patterns known from
laboratory experiments on Rayleigh--B\'enard convection and may be
an imprint of the pattern of subphotospheric convection. It was
also noted that averaging does not completely smear the image,
which still comprises a multitude of granular-sized, light
``blotches'' against a darker background. In some cases, the time
variations of intensity at the point corresponding to the
averaged-intensity maximum in such a blotch and to a nearby
minimum exhibit a tendency to anticorrelation. We interpreted all
these findings as evidence for the presence of a previously
unknown type of self-organization in the solar atmosphere.

After paper I appeared, Rast \cite{rast} disputed our conjecture.
He suggested that the features of granulation patterns reported by
us are merely of statistical nature and do not reflect the
structure of real flows. To substantiate this suggestion, he
artificially constructed a series of random fields with some
characteristic parameters typical of solar granulation. Rast
claimed that the features of regularity described in paper I can
be reproduced even if such artificial fields are used instead of
real photospheric images. On this basis, he denied that our
observations had any physical implications.

Here, a more extensive investigation of the quasi-regular
structures is presented. In addition to the La Palma series, we
consider a 45.5-h series of white-light images obtained with the
SOHO MDI instrument. We analyse movies constructed by taking
running averages on these two series of images, employ some
techniques of algorithmic treatment of the images, and note
remarkable features of spatio-temporal intensity correlation
related to the structures.

Although some doubts about the reality of our elusive subject may
still remain, the results that will be presented here and in a
companion paper to be written in coauthorship with P. N. Brandt
additionally testify to the actual existence of long-lived,
quasi-regular structures. In particular, we can decline the
critical remarks by Rast \cite{rast}.

\section{Observations and Primary Data Reduction}
The La Palma series of photospheric images was obtained by Brandt,
Schar\-mer, and Simon (see Simon {\it et al}., 1994) on 5 June
1993 using the Swedish Vacuum Solar Telescope (La Palma, Canary
Islands). It still remains unsurpassed in its duration (11 h),
continuity (a constant, 21-s frame cadence), and quality (rms
contrast varying between 6 and 10.6\%).

The observations lasted from 08:07 to 19:07 UT on 5 June 1993.
Images of a $118.7\times 87.9$ Mm$^2$ area of the solar
photosphere, not far from the disk centre, were produced by the
telescope in the 10-nm-wide spectral band centred at a wavelength
of 468 nm. Frames were recorded every 21.03 s by a CCD camera with
a pixel size of 0.125$\arcsec$, after selecting each of them as
having the highest contrast among 55 images obtained during the
first 15 s of the 21.03-s cycle. The resolution was typically no
worse than about 0.5$\arcsec$.

The pre-processing of data included the following three principal
steps. First, images were aligned by shifting the next image
relative to any current one so as to maximize the correlation
between them. Second, a destretching procedure based on the
technique of local correlation tracking (November, 1986) was used
to compensate for atmospheric distortions. Third, fast intensity
variations were removed by subsonic Fourier filtering (Title {\it
et al}., 1989) with a cutoff phase speed of 4 km s$^{-1}$. A
Fourier transform technique was applied to interpolate the images
to equal time cadence of 21.03 s. The entire series contains
almost 1900 frames. For a more detailed description of the
data-acquisition technique see Simon {\it et al.} (1994).

For our analyses, we use a subset of the series that covers an
area of $43.5\times 43.5$ Mm$^2$ ($60\arcsec\times 60\arcsec$, or
$480\times 480$ pixels) and an interval of length 8 h 45 min (1500
frames). The contrast of the printed images in this paper is
artificially enhanced.

In addition to the La Palma series, we analysed the 45.5-h series
obtained with the SOHO MDI instrument in 1997, from 17 January
00:01 UT to 18 January 21:30 UT (see Shine {\it et al.}, 2000).
This series contains white-light images with a resolution of about
1.2$\arcsec$\ taken at a 1-min interval.

\begin{figure} 
 \centering
 {\includegraphics[bb=3 0 340 266,clip,
 width=9cm]
 {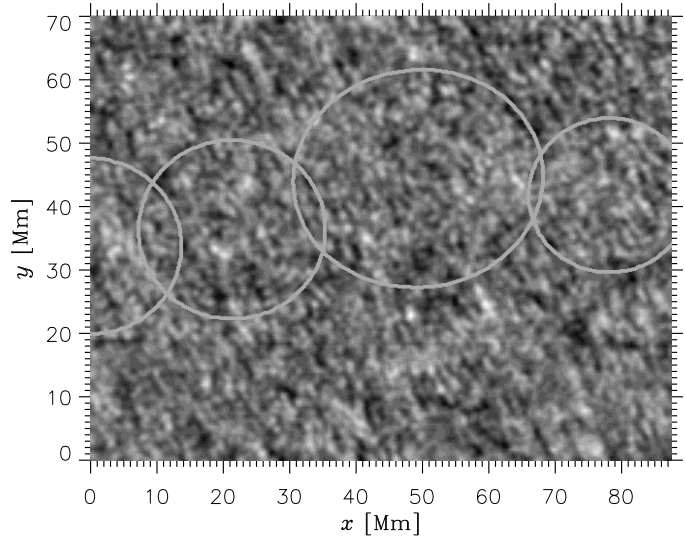}}
\caption{%
A sample of 2-h-averaged images of the SOHO MDI series with a
pronounced trenching pattern. The light grey ellipse and circles
mark concentric-ring structures. Multiple nearly-straight trenches
and ridges form overall ``hatching'' diagonally downward to the
right. }\label{MDI}\end{figure}

We mainly dealt with a subsonically filtered version of this
series, which is free of five-minute oscillations and whose frames
are fixed to a certain location on the solar surface. The filtered
images are $304\times 480$ pixels in size, a pixel being about
0.6$\arcsec$ large. Accordingly, the area covered by an image
measures $182\arcsec\times 288\arcsec$, or about $132 \times 209$
Mm$^2$. Trenching patterns are especially pronounced in enlarged
fragments of averaged images. For example, Figure \ref{MDI}
represents a cutout that contains $200 \times 160$ pixels and
measures about $87 \times 70$ Mm$^2$.

\section{Movies}
The very idea of implementing a study of well-correlated
structures on the solar surface was suggested to us by a close
examination of the movie representing the dynamics of granulation
patterns that was obtained by Title {\it et al.} \cite{title}
using the SOAP instrument of the \emph{Spacelab 2} optical
laboratory on the \emph{Challenger} space shuttle. If such a movie
is viewed at a sufficiently low rate and especially in a
back-and-forth playback mode over short intervals, signs of
regularity become quite visible: the proper motions of granules
appear to be organized in roll motions. While the roll widths are
somewhat larger than the characteristic size of granules, the
rolls are stretched over fairly long distances and, in some cases,
form closed rings.

Generally, movies are very convenient for the visual
identification of characteristic features of evolving patterns.
Since time-averaged images that may visualize long-lived
photospheric structures are a subject of our particular interest,
we constructed movies of running-average sequences of images. In
other words, each frame of such a movie is an image averaged over
the same interval, while the central time of the averaging
interval changes by a fixed increment between contiguous frames.
If the averaging time is properly chosen, a careful inspection of
running-average movies can enable us to detect features of
interest and to follow their evolution. At the moment, we regard
averaging times of about 1--2 h to be nearly optimal for the
detection of quasi-regular structures. Mainly, we deal with 2-h
running averages.

Our examination of running-average movies constructed from both
the La Palma and SOHO MDI series has revealed numerous
quasi-regular patterns, so that the presence of ``trenching''
patterns in the distributions of the time-averaged brightness no
longer appears to be an unusual phenomenon. In particular, the
distribution of light blotches and dark gaps between them is in
many cases remarkably anisotropic. An example of a highly trenched
pattern is given in Figure \ref{MDI}, where numerous ridges and
trenches forming rightward downward ``hatching'' can be seen. In
this case, the anisotropy in the distribution of linear chains of
blotches is obvious (the averaging time is here 2 h). In addition,
families of nearly concentric, not quite regular rings can be
distinguished. They are marked with a light grey ellipse and light
grey circles in the figure.

We emphasize that, in contrast to what Rast \cite{rast} claims,
his artificial random fields contain upon averaging only isolated
linear features rather than families of such features and do not
resemble averaged solar images.

\begin{figure}[!t] 
\centering
 \includegraphics[bb=0 0 340 340pt,
 width=7cm]{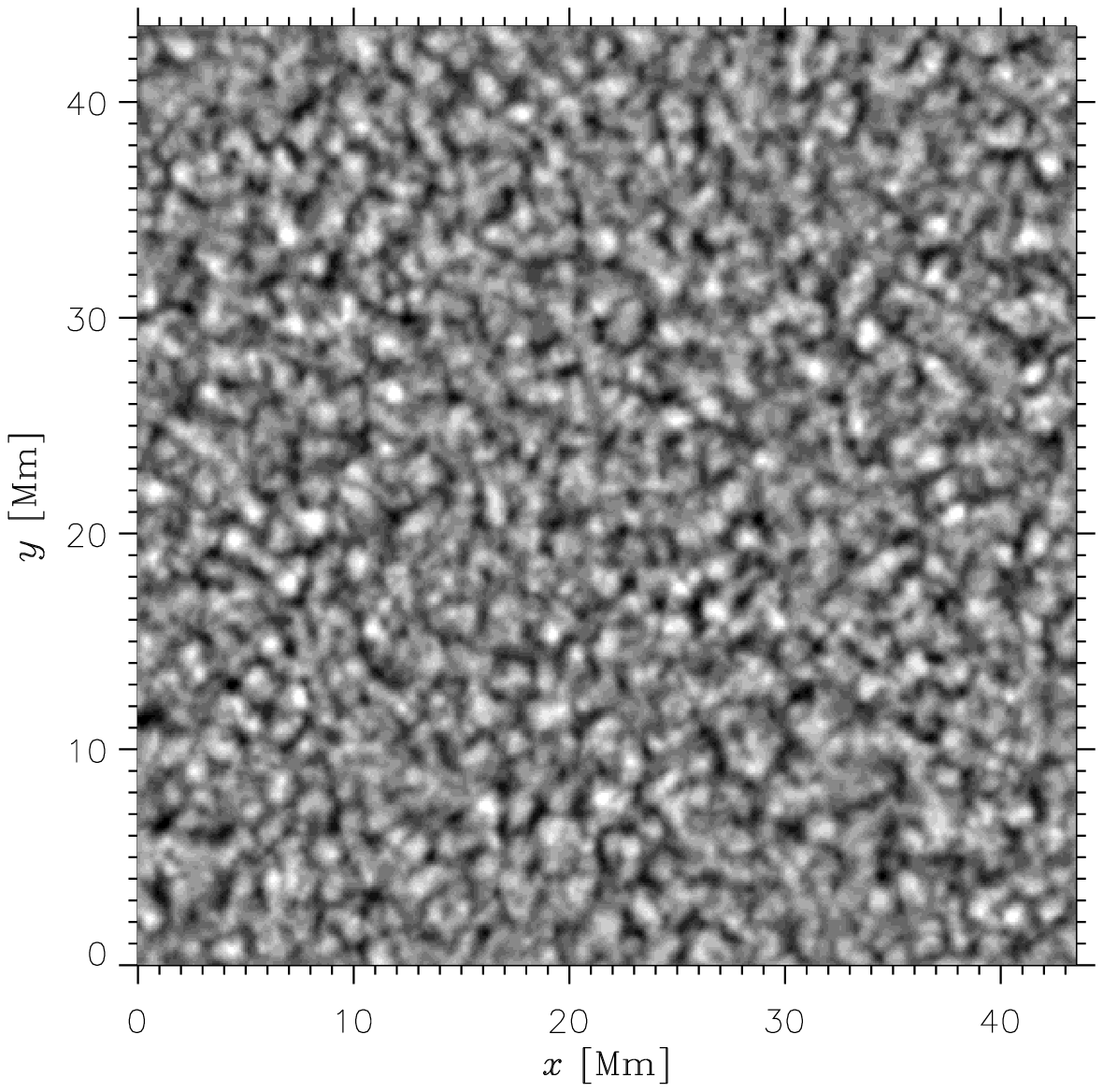}
 \includegraphics[bb=0 0 340 340pt,
 width=7cm]{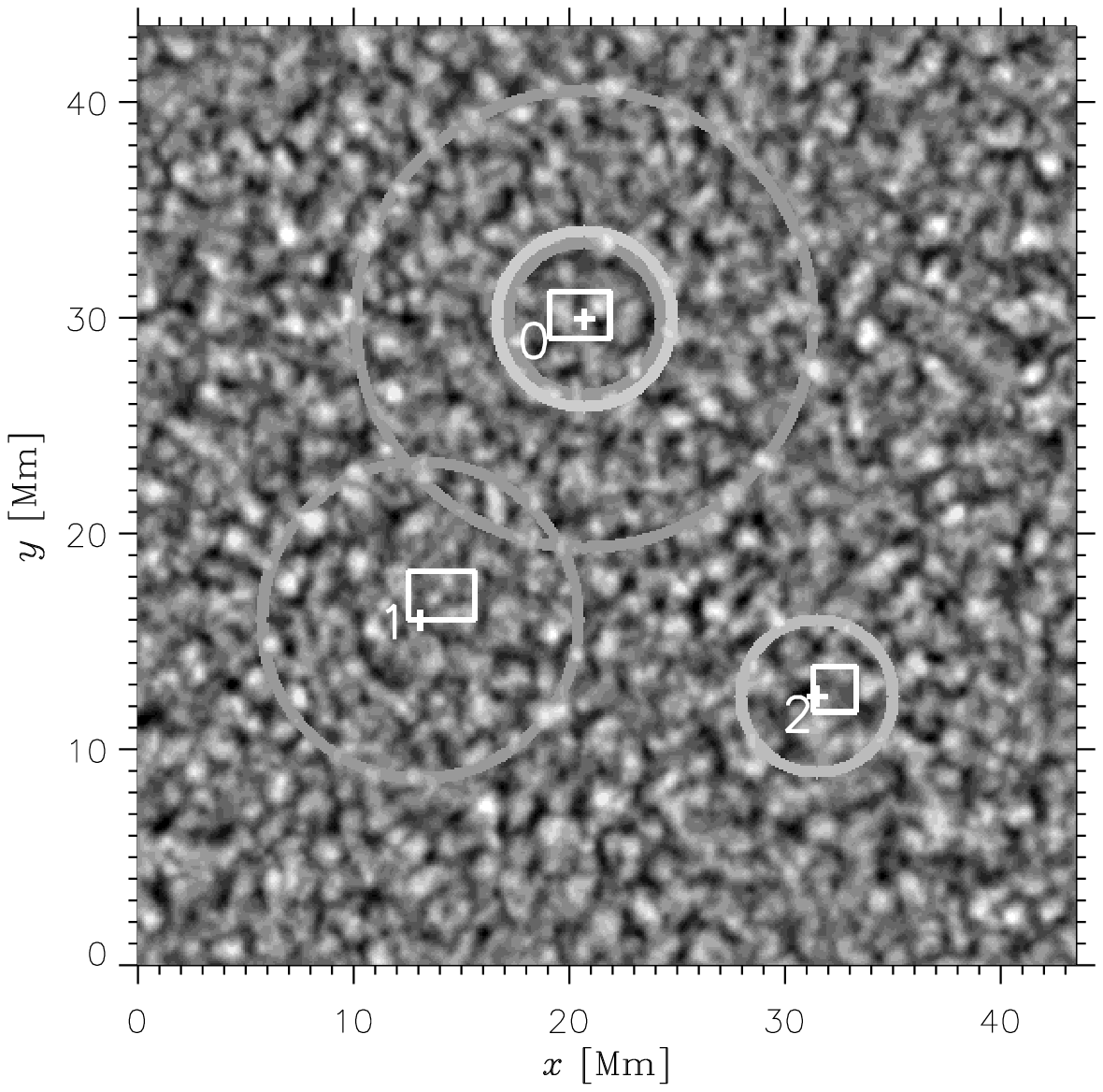}
\caption{%
Top: A sample of 2-h-averaged images of the La Palma series.
Bottom: The same image superposed with patterns of azimuthally
averaged intensity for the most likely positions of the centres of
ring systems (crosses) detected by our algorithm. The scanned
areas are marked with white rectangular frames and numbered. A
common intensity scale is used for all centres. A web pattern can
be distinguished in the original image --- its centre is located
within the area labelled as 0 in the lower panel.
}\label{spokes}\end{figure}

Families of concentric rings are sometimes superposed with radial
``spokes'', so that ``web patterns'' can be observed (see Figure
\ref{spokes}). Ring and web patterns admit a hydrodynamic
interpretation in terms of the development of certain
instabilities of a larger-scale upwelling that can be associated
with meso- or supergranules. We plan to consider this point
elsewhere.

\begin{figure}[!p] 
\centering
 \includegraphics[bb=0 0 340 340pt,
 width=7cm]{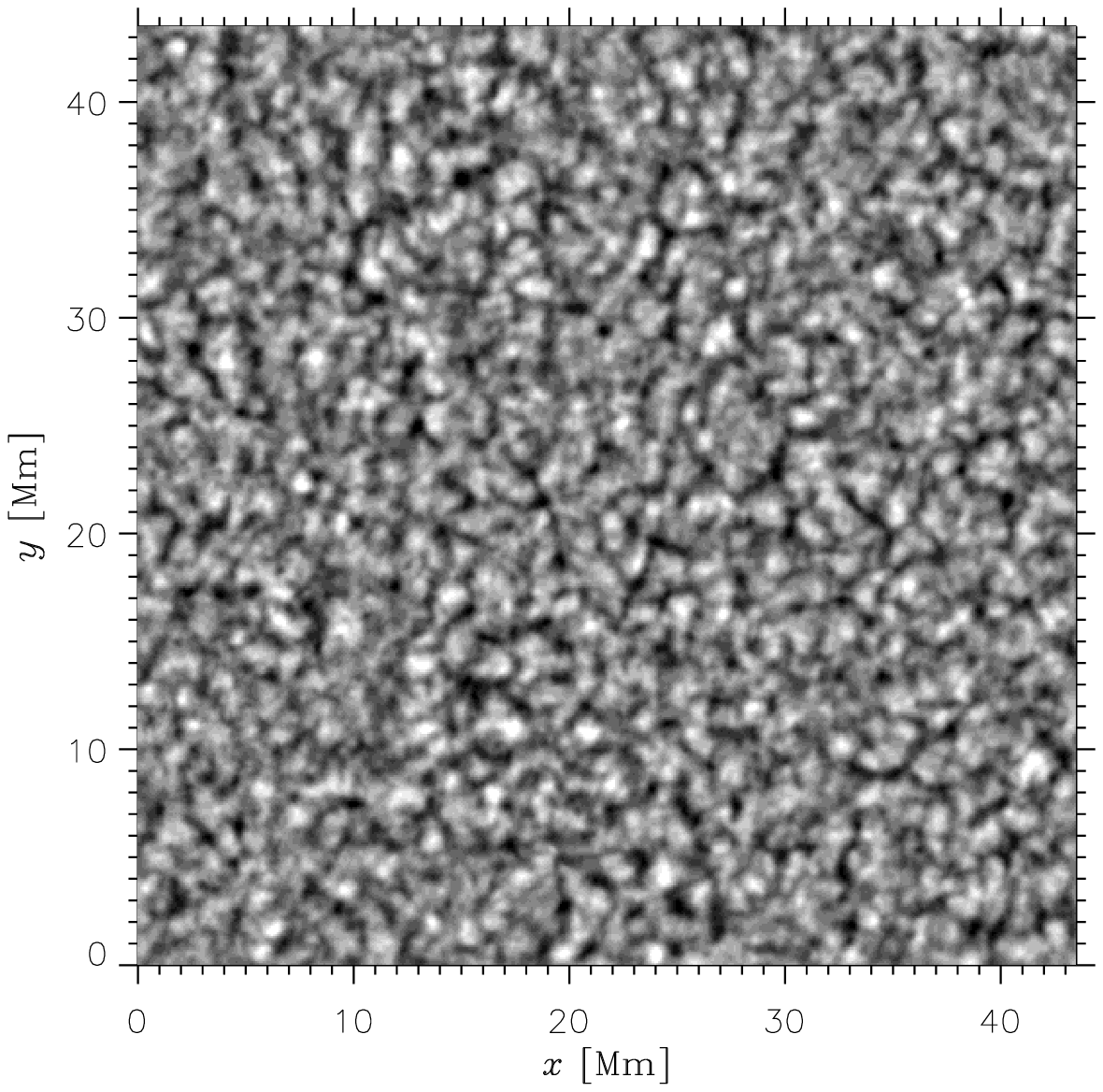}
 \includegraphics[bb=0 0 340 340pt,
 width=7cm]{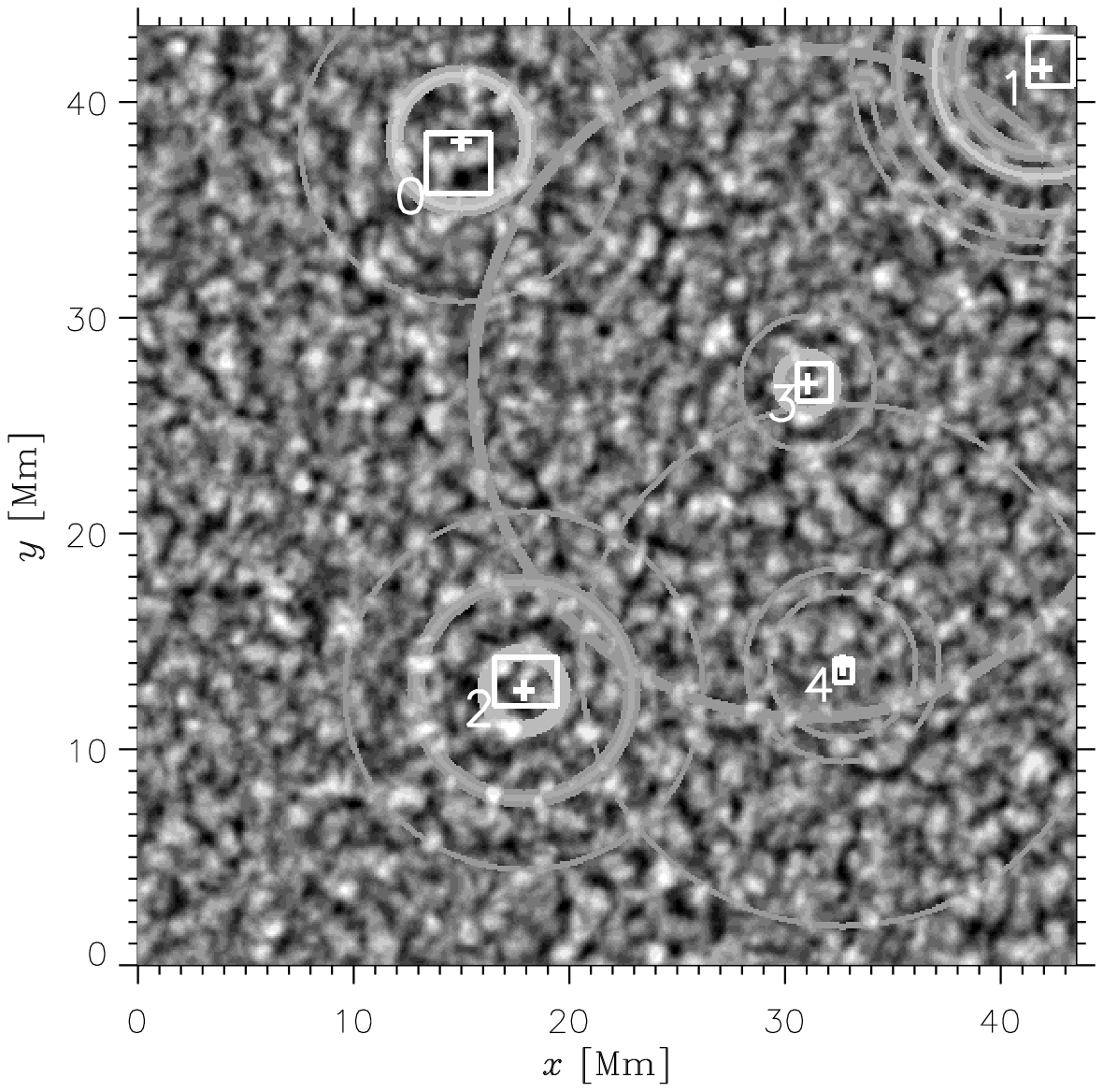}
 \includegraphics[bb=0 0 340 340pt,
 width=7cm]{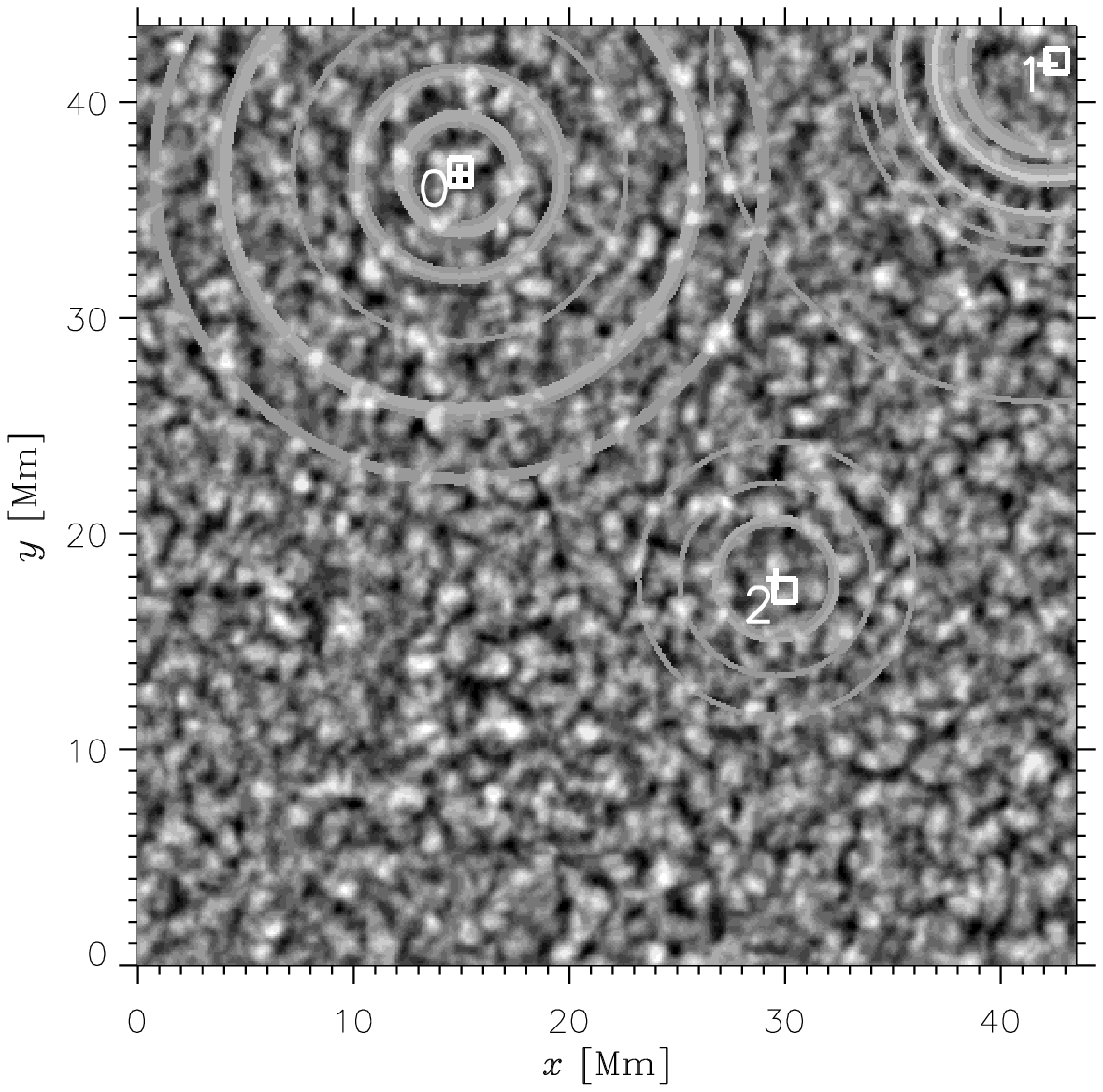}
\caption{%
Top: A sample of 2-h-averaged images of the La Palma series.
Middle and bottom: The same image superposed with patterns of
azimuthally averaged intensity for the most likely positions of
the centres of ring systems (crosses) detected by our algorithm.
The scanned areas are marked with white rectangular frames and
numbered. A common intensity scale is used for all centres in each
panel. }\label{5areas}\end{figure}

\begin{figure}[!t] 
\centering
 \includegraphics[bb=0 7 481 283pt,
 width=7.2cm]{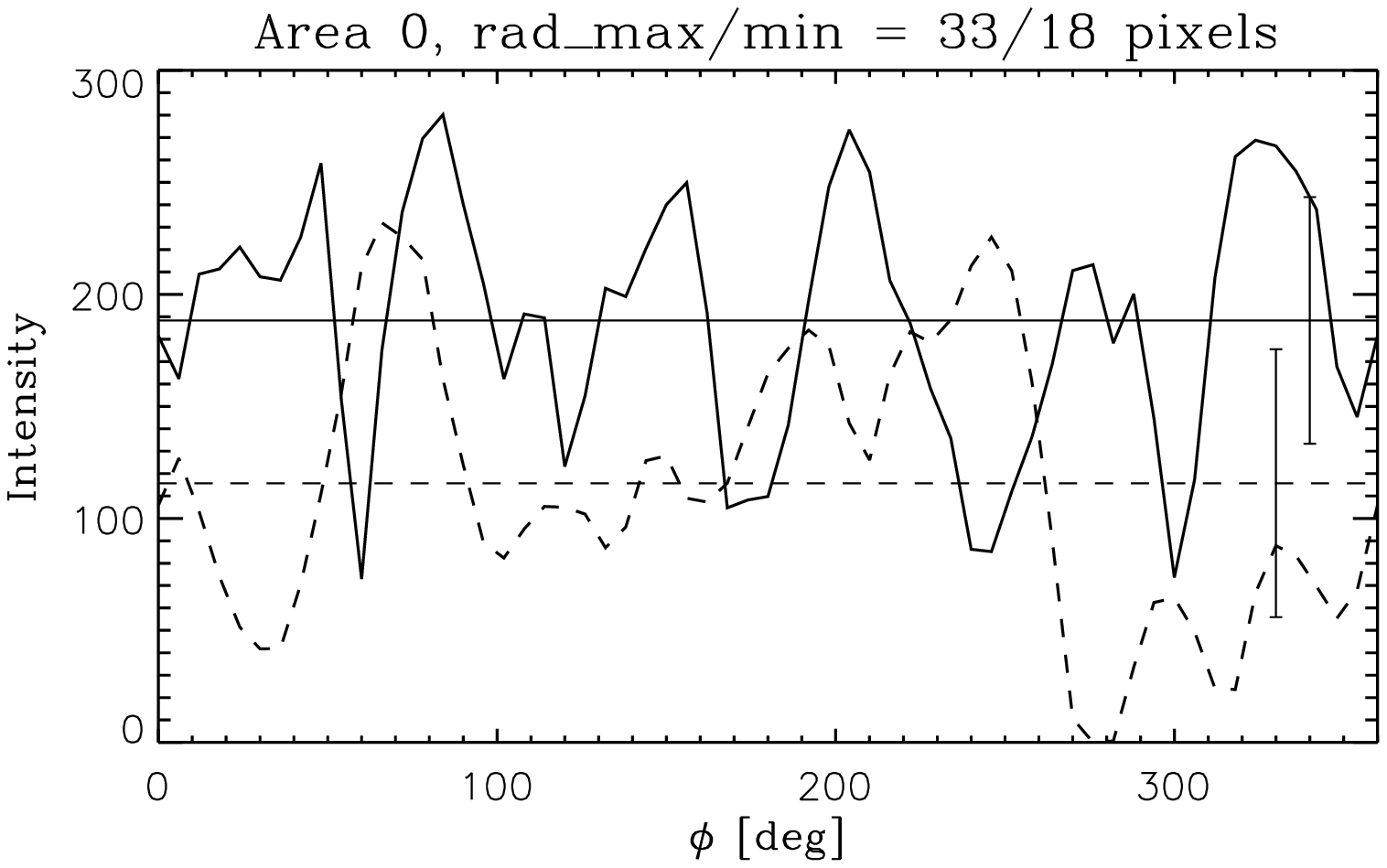}
\includegraphics[bb=0 7 481 283pt,
 width=7.2cm]{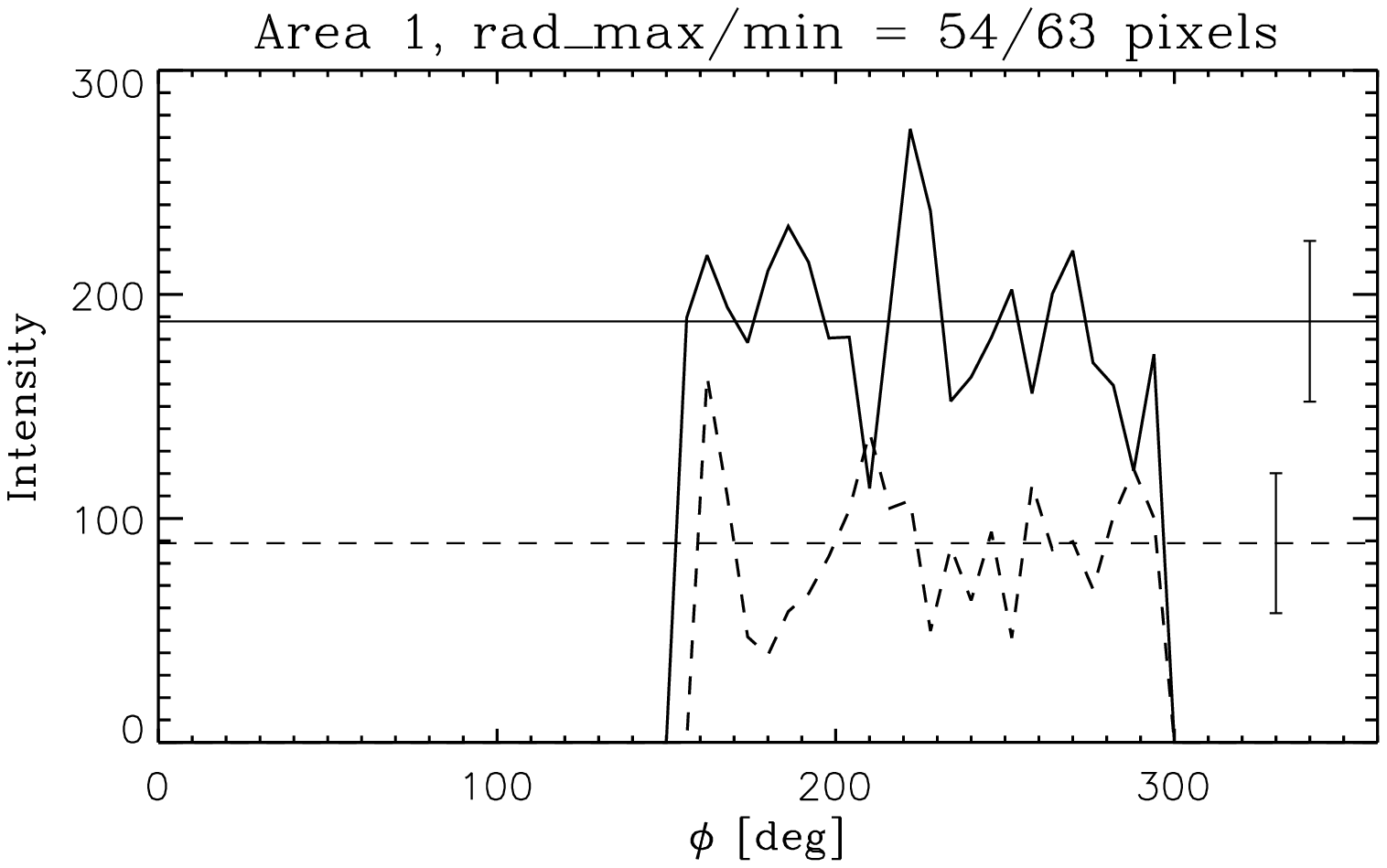}
\includegraphics[bb=0 0 481 283pt,
 width=7.2cm]{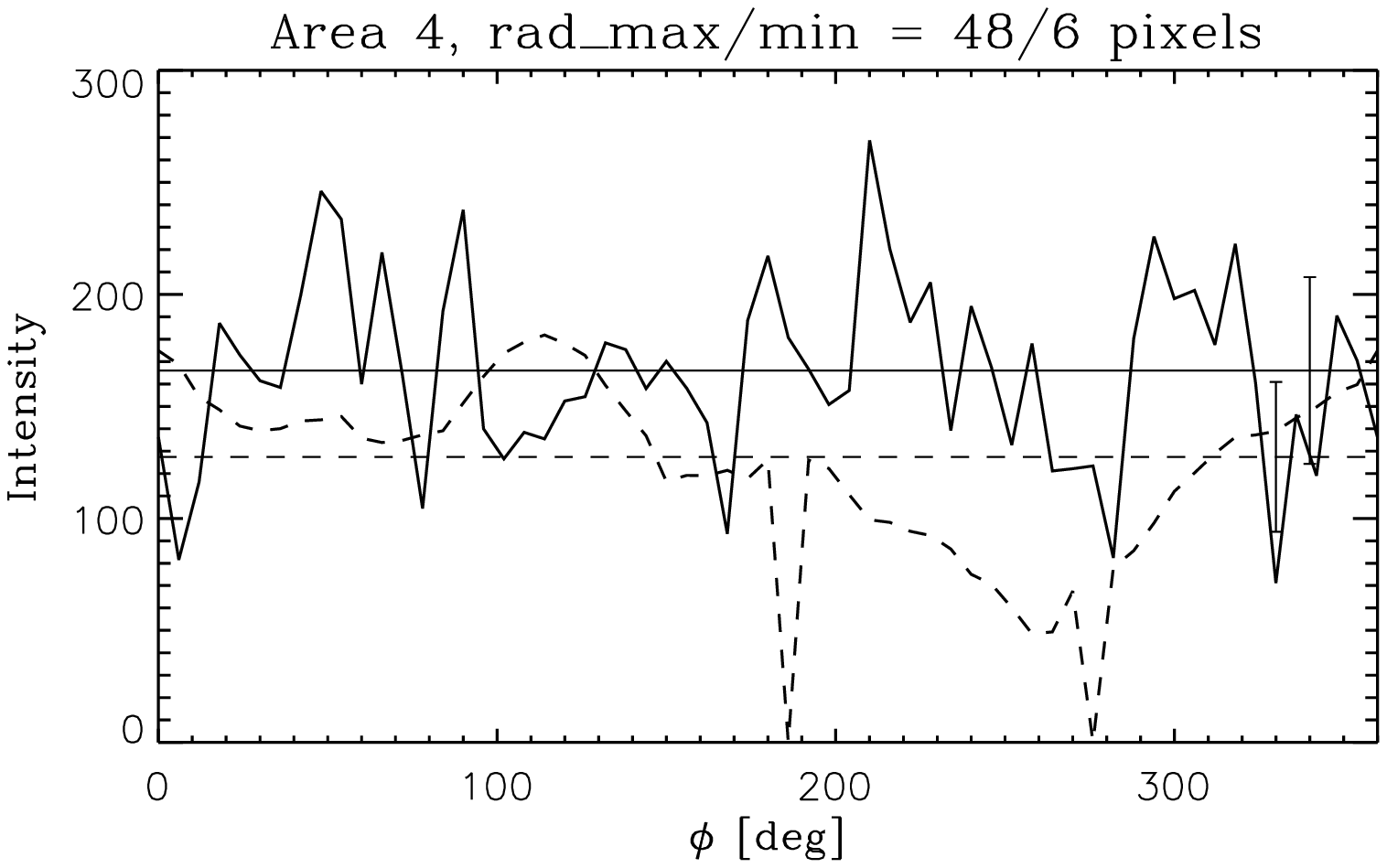}
\caption{%
Intensity as a function of the azimuthal angle $\phi$ at the
radial distances where the azimuthally averaged intensity reaches
its maximum (solid curve) and minimum (dashed curve), for three of
the centres marked in Figure \ref{5areas}, middle. The area number
and these radial distances (in pixels) are indicated at the upper
edge of each panel. The horizontal lines correspond to the
azimuthally averaged intensities (solid for the maximum and dashed
for the minimum) and the vertical bars to the standard deviations.
}\label{azvar}\end{figure}

\section{Algorithmic Treatment of Averaged Images}
Our principal difficulty, which severely restricts possibilities
of algorithmic detection of quasi-regular structures, stems from
the very low signal-to-noise ratios. By the noise, we mean here
the chaotic, blotchy background in which the quasi-regular
patterns --- the subject of our study --- are ``dissolved''. Up to
now, none of the tested algorithms has been found to be
universally capable of filtering out such a noise and detecting
the ordered component of the averaged-intensity field.
Nonetheless, some noteworthy features can be revealed using our
algorithm described below. A possible alternative approach is
based on using a software package intended for the detection of
geological structures in aerospace images.

\subsection{Analyses of azimuthally averaged brightness distributions}
Our algorithm scans given rectangular areas in the image, uses
each point as a trial centre, computes the radial distributions of
the azimuthally averaged intensity, and plots rings at the local
maxima of the averaged intensity.

Before constructing a time average, we normalize each individual
image by setting the mean intensity to a certain level, universal
to all images, and remove the residual large-scale intensity
gradients. Within the averaged image, we delimit one or more
rectangular areas in which the program will seek the most
plausible position of the centre of a ring system. We also specify
the maximum ring radius and the radial and azimuthal sizes of the
bins that will be used to compute the brightness distribution.

Each area is scanned over two Cartesian coordinates, and each
trial point (pixel) is regarded as the origin of an azimuthal
coordinate system. The radial coordinate in this system is divided
into small intervals --- radial bins --- and the intensity is
averaged over each narrow annulus corresponding to a radial bin.
This yields the radial distribution of the azimuthally averaged
intensity referenced to the trial centre. If the centre lies in a
light blotch, small-radius annuli may completely fall within this
blotch, so that the averages over these annuli may be spuriously
large. To avoid overestimating the role of the central pixels, the
averaged intensities between the centre and the first local
minimum next to the first local maximum are set equal to this
local minimum. Among all radial distributions computed for the
given area, the algorithm selects the one that exhibits the widest
range of radial variation in the azimuthally averaged intensity.
The corresponding trial centre is considered the best candidate
for being the centre of some ring system.

If more than one area is selected in a given averaged image, the
algorithm finds such a centre in each area and applies a common
normalization to all radial distributions obtained for these
centres. The ring pattern is visualized as follows. The absolute
minimum of all distributions selected in the given image is
subtracted from these distributions. In all of them, the radially
averaged intensity is then set to zero wherever it does not exceed
a chosen threshold value, and the resulting ``apodized''
distributions are additively superposed onto the original image.
Except the two-dimensional maps of ring patterns thus obtained,
two azimuthal distributions of the true (local) intensity are
plotted for each centre; they correspond to the radial distances
at which the azimuthally averaged intensities are maximum and
minimum.

Figure \ref{5areas} presents some results so obtained for one of
the best available (in terms of the discernibility of structures)
time-averaged images of the La Palma series. The image itself is
shown in the upper panel. In two other panels, this image is
superposed with the distributions of the azimuthally averaged
intensity for some centres selected by the algorithm. The areas
scanned by the algorithm are marked by white rectangles and
numbered.

For some centres marked in the middle panel, Figure \ref{azvar}
shows the local intensity as a function of the azimuthal angle
$\phi$, the azimuthally averaged intensity, and the standard
deviations at the radii where the azimuthal averages are maximum
and minimum. We see that the centres of clear-cut ring systems are
characterized by especially large amplitudes of radial variations
in the azimuthally averaged intensity. These are the centres
marked in areas 0 and 1; the curves for area 2 in the middle panel
and areas 0 and 1 in the bottom panel (not presented here) behave
similarly. In contrast, the lowermost graph in Figure \ref{azvar},
shown for comparison, clearly illustrates the noisy character of
the pattern around the centre found in area 4. This is consistent
with the fact that no pronounced ring system has been detected for
this area.

In both the middle and bottom panels of Figure \ref{5areas}, area
0 is associated with the same, most pronounced, ring system. The
reasons for the dramatic difference in the corresponding radial
distributions of the azimuthally averaged intensity are as
follows. In the first case, the trial rectangular area was
considerably larger, chosen in the hope that the formal criterion
of maximum radial-variation range would be sufficient for the
detection of a well-developed ring system. In the second case, the
choice of the area was aimed at capturing the expected
(``resonant'') centre of the visible ring system, while the centre
selected in the first case fell outside this area; thus, the
algorithm detected an intensity distribution outlining the visible
ring pattern more clearly, although the range of radial variations
in the azimuthally averaged intensity was somewhat smaller than in
the first case. Obviously, the noise level is so high that it can
accidentally yield a wider range for a centre other than the
centre of the really present ring system. If, however, the
algorithm, scanning the chosen area, passes through the resonant
position of the trial centre, it detects the ring system quite
successfully.

Likewise, area 1 is chosen in both cases so as to detect the
pronounced ring system near the upper right corner of the
considered region. In the bottom panel, area 1 is again smaller
than in the middle panel and does not include the centre obtained
using the larger area and marked in the middle panel. It is
noteworthy, however, that the algorithm nevertheless detects the
ring pattern that has nearly the same appearance whichever of
these two trial areas is used. As can be seen from the graph for
area 1 in Figure \ref{azvar}, a large difference between the
maximum and minimum intensities is characteristic of this ring
pattern; at the same time, the standard deviations are especially
small, which is indicative of a relatively low noise level in this
region.

Another example of applying our algorithm to time-averaged
granulation images is given in Figure \ref{spokes}. Here, the
centre detected in area 0 corresponds to a system that includes a
pronounced ring about 8 Mm across, a much fainter ring (hardly
discernible by eye) about 21 Mm across, and radial ``spokes''
within the smaller ring. On the contrary, the ring centred in area
1, which is very faint and isolated, simply represents the maximum
of the radially averaged noise intensity and seems to have no
physical meaning (such a maximum is necessarily plotted for any
area provided the maximum intensity measured in units of the
absolute maximum for all selected areas, exceeds some
user-specified threshold). Apparently, this also applies to area
2, where the plotted ring is likewise isolated and faint.

Thus, although this algorithm is fairly sensitive to the level of
noise in the analysed intensity field, not only may it be
illustrative of qualitative differences between ring systems and
purely chaotic patterns but it can also be used, with due care, to
test hypotheses of the presence of faint ring systems. Further
improvements of the identification criterion for ring patterns
could enhance the potentialities of the algorithm.

A limitation of our algorithm lies in the fact that it assumes the
structures to be strictly circular and cannot adapt itself to
real, imperfect circles. To extend the scope for algorithmic
treatment of structures in the solar granulation, attempts were
made to apply a different approach.

\begin{figure}[!t] 
 \centerline{
 \includegraphics[width=6.3cm,bb=8 0 340 340pt]
  {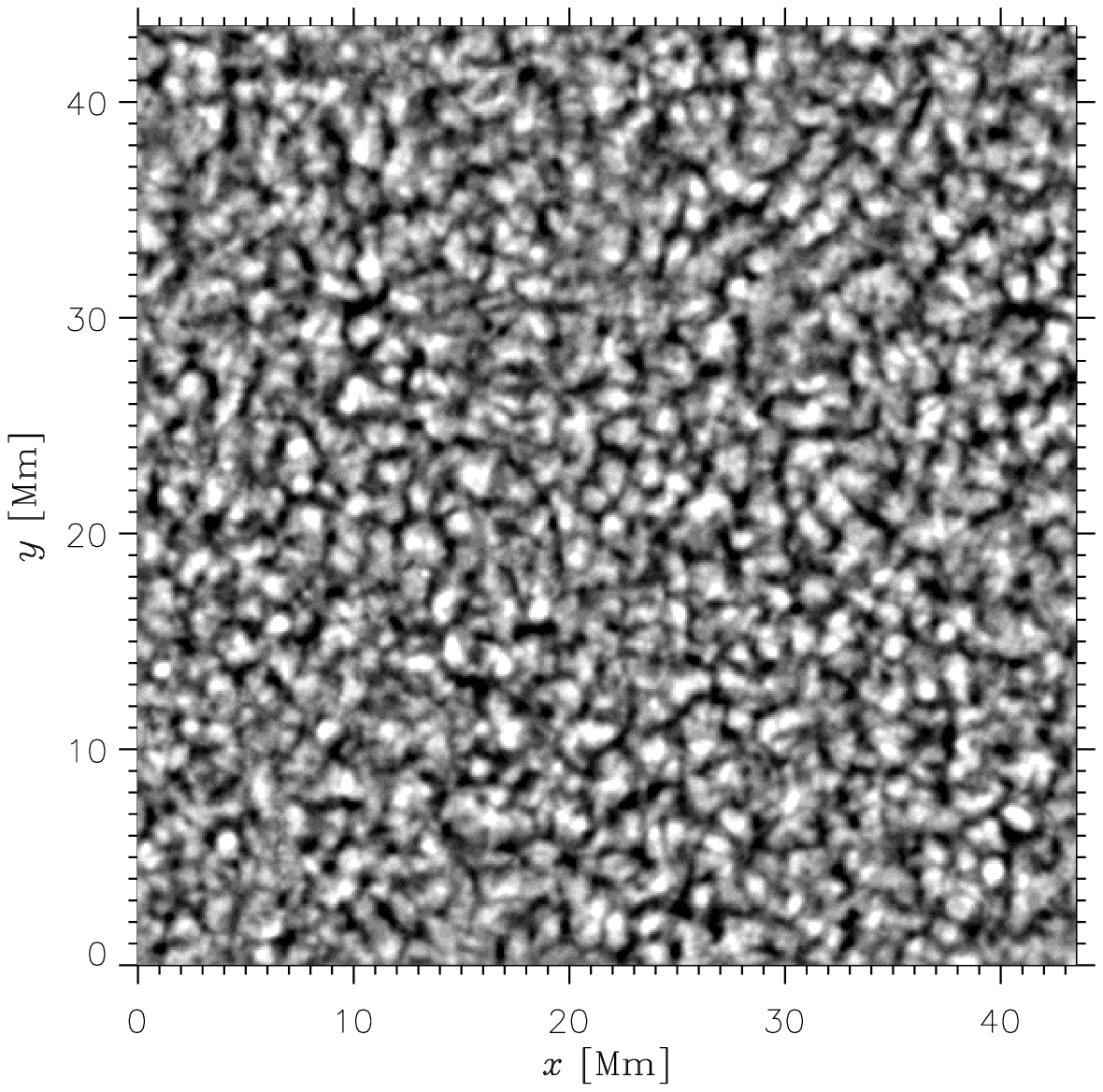}
 \includegraphics[width=6.0723cm,bb=20 0 340 340pt,clip]
  {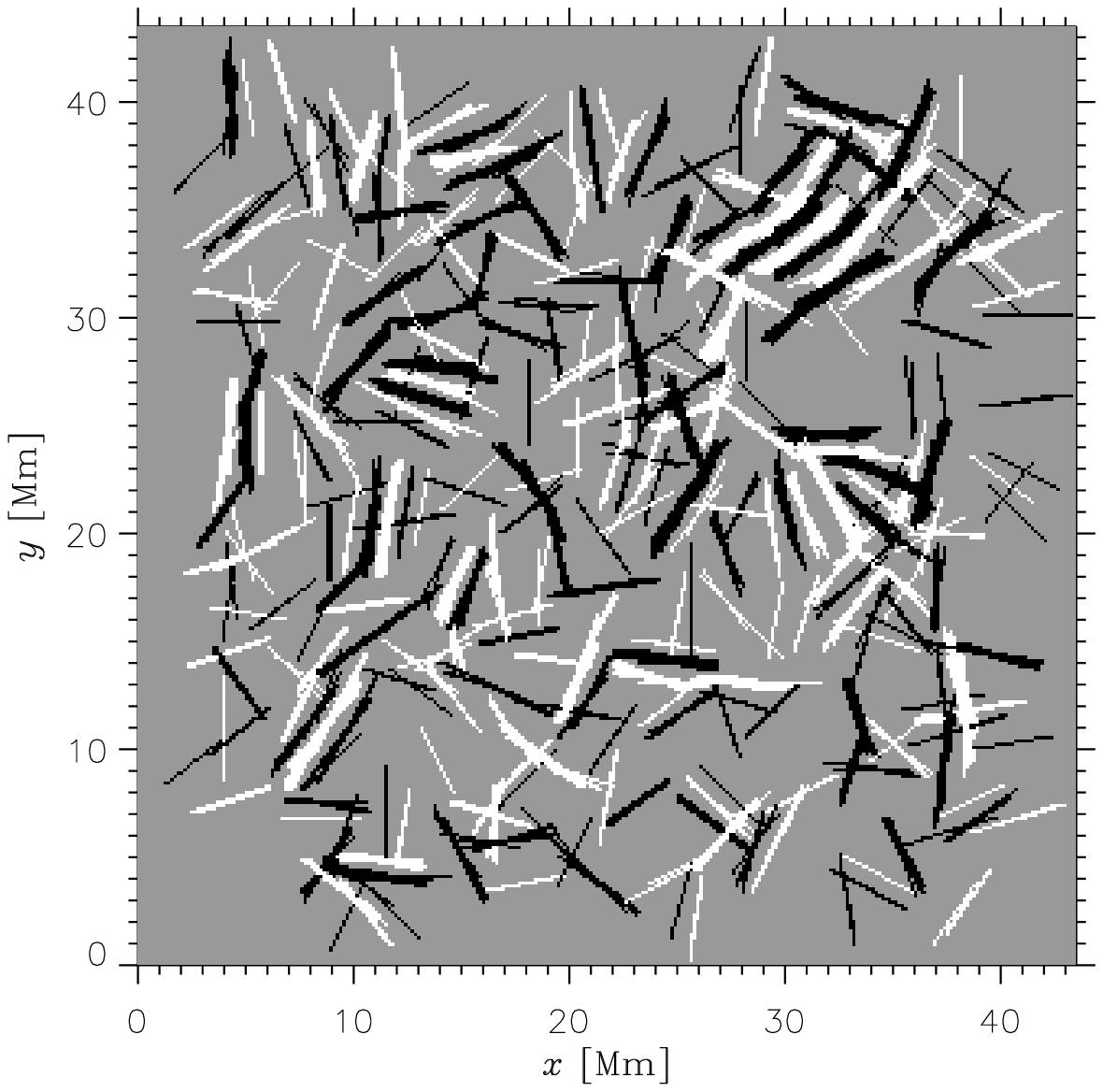}}
\caption{%
Processing of a sample of 1-h-averaged images of the La Palma
series carried out by A.A. Buchnev using Salov's algorithm. Left;
original image: Right; detected light and dark linear structures
on a grey background. }\label{lineam}\end{figure}

\subsection{Application of an algorithm intended for aerospace-image processing}
A few representative time-averaged images were tentatively
analysed by A.A. Buchnev (Institute of Computational Mathematics
and Mathematical Geophysics, Novosibirsk, Russia) using the
software package developed at the same institute by Salov
\cite{salov} to detect linear and circular geological structures
in aerospace images of the Earth's surface (see also Buchnev {\it
et al.}, 1999). The algorithm that detects such structures uses a
nonparametric statistical criterion, seeking lines in the image
along which the intensity values are systematically higher or
lower than the values at points located symmetrically on both
sides of the line.

Figure \ref{lineam} exemplifies the results of such a procedure.
The program was run in the mode of seeking linear structures
(lineaments). Curved features could generally be described as
chains of lineaments of varying length. Both light and dark
lineaments were detected, and their graphic representation is most
descriptive if they are plotted on a grey background (Figure
\ref{lineam}, right).

The most remarkable feature in this map of lineaments is a
pronounced trenching pattern that includes families of parallel,
alternating light and dark lineaments and chains of lineaments. If
we ignore some purely disordered fragments of this pattern, we can
note that, in some cases, lineaments forming different patches of
this pattern can be connected by means of continuing them over the
gaps between the patches, so that fairly extended families of
lineaments can be detected. In particular, one of such families
runs over the upper right quadrant of the frame, from the corner
to the central region. It is to such objects that a physical
meaning should be attributed in the first instance. Let us also
note that most part of the lower left quadrant is encircled with a
nearly elliptic contour intersected by a family of chains formed
by parallel light and dark lineaments.

This example demonstrates that, although much care must be taken
in using this image-processing procedure, the results obtained
here show promise for the detection of hidden trenching patterns
in time-averaged granulation images.

\begin{figure}[!t] 
\centering
 \includegraphics[width=7.5cm,
 bb=15 10 453 260,clip]
  {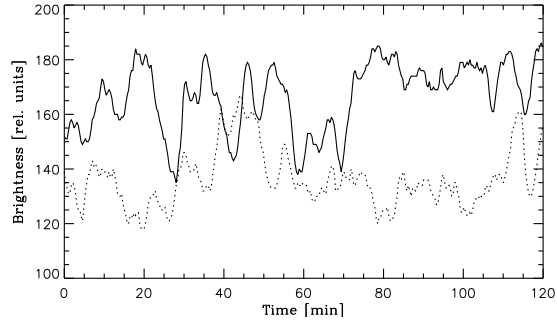}\\
\caption{%
Pair of brightness-variation curves for the points of a local
intensity maximum (solid curve) and a nearby minimum (dotted
curve) chosen in a 2-h-averaged solar image.
}\label{brvar}\end{figure}

\section{Running-Average-Based Correlations}
The strong brightness nonuniformity of the averaged granulation
images (the presence of light and dark blotches) suggests that the
probability of emergence of granules is likewise nonuniformly
distributed over the solar surface, and granules ``prefer''
certain locations. From the standpoint of comprehending the
physics of the phenomenon, it seems worthwhile to investigate the
relationships between granule-emergence events at various points.

As thermal convection takes place in a horizontal fluid layer
heated from below, the horizontal temperature distribution at a
given height reproduces, to a first approximation, the horizontal
distribution of the vertical velocity component at the same
height. According to the viewpoint we hold (Getling, 2000), which
agrees with some other analyses (Rieutord {\it et al.}, 2001) and
which will be further substantiated below, granules --- blobs of
overheated material\footnote{Even in classical Rayleigh--B\'enard
convection with the simplest, purely conductive energy-transfer
law, such blobs can develop as the product of some instability
mode ---the one-blob or two-blob instability (Bolton and Busse,
1986). Under solar conditions, radiative transfer can additionally
destabilize the process, since it enhances the effect of various
overheating instabilities.} --- are relatively passive tracers of
larger-scale convective motions, associated nevertheless with the
small-scale ``noise'' in the velocity field. In this case, the
distributions of temperature and vertical velocity over the
photospheric surface on a mesogranular scale and up should be
similar only when averaged over time. Likewise, we can expect the
time-averaged field of the vertical velocity component to be
mainly reflected by the time-averaged brightness field. Thus,
steady convective upflows and downflows will appear light and dark
in time-averaged images, respectively.

In paper I, we noted a remarkable property of the time variations
of brightness at two points chosen in a 2-h-averaged granulation
image (corresponding to nearly the same averaging interval as in
the case of our Figure \ref{5areas}). One of these points
corresponded to a local intensity maximum (which is in Figure
\ref{5areas} among the light blotches forming one of the light
rings centred at point 0) and the other to a nearby minimum. The
curves of brightness variation (not presented in paper I) are
shown here in Figure \ref{brvar}. It can be seen that the
intensity values at the two points exhibit apparent
anticorrelation.

To check this immediate impression and in view of forming an idea
of the dynamics of subphotospheric flows, it seems useful to
calculate correlations between time variations of brightness at
various points of the granular field. Assume that isolated hot
blobs are present in the circulating material. Let the intensities
at two points of the solar surface, presumably located at nearly
the same streamline, be $I_1(t)$ and $I_2(t)$. Then some time
scale $\tau$ of the order of the convective-circulation period
should manifest itself in both these intensity variations. If
there is no interference of other processes affecting the
brightness at the considered location, $I_1(t)$ and $I_2(t)$
should correlate fairly well over a time interval $T$
corresponding to the lifetime of the local circulation system. If,
however, such interference is present, the pattern of correlations
may be substantially complicated and eventually smeared.

On the other hand, interference on a time scale $\mathcal T$
considerably longer than $\tau$, which would by itself produce an
``interfering'' correlation, can be eliminated as follows. Assume
that, when computing the correlation, instead of the normal
average values of $I_1(t)$ and $I_2(t)$ calculated over the whole
time interval considered, we use their running averages obtained
with a window wide enough to smooth out short-term variations on
the time scale $\tau$ but narrow enough to retain the long-term
variations on time scales comparable with $\mathcal T$. In this
case, the fluctuations of $I_1(t)$ and $I_2(t)$ will be measured
from the levels of the smoothed values, which experience the
long-term variations, and these variations will not be taken into
account in the resulting correlations. Thus, we shall obtain the
correlation between the variations on the time scale $\tau$, while
the effect of variations on the time scale $\mathcal T$ will be
filtered out.

Let us consider the intensity variations at two points that are
located not far apart and appear light and dark in the
time-averaged image. Such points may prove to be the places of
emergence and sinking of a hot blob. The lag between the
brightness variations at the two points will correspond to the
time taken by the blob to traverse the distance between these
points. If the characteristic lifetime of the blob is larger than
the circulation period, the correlation curve may exhibit
additional correlation peaks separated from the main peak by this
period (and, generally, its multiples). A moving temperature
minimum at the same streamline, associated with cool material,
will produce a negative extremum of the correlation function at a
lag corresponding to the time interval between the passage of the
hot blob through one point and the cool blob through the other.

To select correlations indicating that the two points may belong
to a common circulation system, we compute running-average-based
cross correlations between local intensity variations as follows.
Let $x_j=(x_0,...,x_{N-1})$ and $y_j=(y_0,...,y_{N-1})$ be two
data arrays with elements corresponding to $N$ instants of time
$t_j=j\Delta t$, $j=0,...,N-1$. If we introduce a moving window of
halfwidth $n\Delta t$, the running average of $x_j$ will be $$
\overline
x_{i,n}=\left\{\begin{array}{lll}\frac{\textstyle{\sum\limits_
{j=i-n}^{i+n}}x_j}{\textstyle \rule{0pt}{8pt}2n+1}&&\mbox{for}\
n\leqslant i\leqslant N\!-\!1\!-\!n,\\[10pt]\overline
x_{n,n}&&\mbox{for}\ 0\leqslant i< n,\\[10pt]\overline
x_{i,n}=\overline x_{N-1-n,n}&&\mbox{for}\
N\!-\!1\!-\!n<i\leqslant N\!-\!1 \end{array}\right.$$ and
similarly for $y_j$. We shall consider the running-average-based
correlations between $x_j$ and $y_j$ defined as
$$P_{xy,n}(L)=\left\{ \begin{array}{lll}\frac
{\textstyle\sum\limits_{k=0}^{N-|L|-1}(x_{k+|L|}-\overline
x_{k+|L|,n})(y_k-\overline
y_{k,n})}{\textstyle\sqrt{\sum\limits_{k=0}^{N-1}(x_k-\overline
x_{k,n})^2\sum \limits_{k=0}^{N-1}(y_k-\overline y_{k,n})^2}}&&
\mbox{for}\ L<0, \\[35pt]
\frac{\textstyle\sum\limits_{k=0}^{N-L-1}(x_k-\overline
x_{k,n})(y_{k+L}-\overline
y_{k+L,n})}{\textstyle\sqrt{\sum\limits_{k=0}^{N-1}(x_k-\overline
x_{k,n})^2\sum \limits_{k=0}^{N-1}(y_k-\overline y_{k,n})^2}}&&
\mbox{for}\ L\geqslant0. \end{array}\right.$$ Obviously, for any
$i$ and $n$, the substitution $$\overline x_{i,n}\rightarrow
\overline x\equiv\frac{\sum\limits_{j=0}^{N-1}x_j}{N},
\quad\overline y_{i,n}\rightarrow \overline
y\equiv\frac{\sum\limits_{j=0}^{N-1}y_j}{N} $$ (which sets the
full width of the window equal to the length of the sample)
reduces the running-average-based cross correlations to cross
correlations defined in a standard way.

\begin{figure}[!t] 
\centering
   \includegraphics[bb=0 4 425 731pt,
   width=7.5cm]{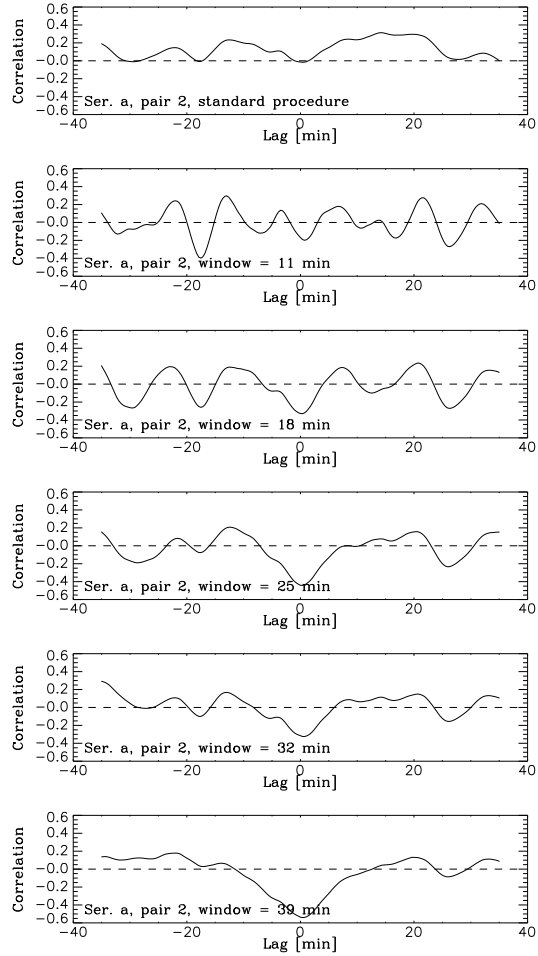}
\caption{%
Correlation curves obtained for a pair of points in the ring
system centred at point 0 (Figure \ref{5areas}) using the standard
technique (top) and running windows of various widths.
}\label{ccs_varwin_a_2}\end{figure}

We present here three characteristic examples of correlations
computed using the running-average technique. All of them refer to
some ``light''--``dark'' pairs of points chosen in the
concentric-ring system centred at point 0 in Figure \ref{5areas}.
In each case, the light point corresponds to a local intensity
maximum on a ridge in the 2-h-averaged image and the dark point is
a nearby minimum in a trench next to the ridge. The correlation
curves thus obtained admit a fairly definite physical
interpretation. Figure \ref{ccs_varwin_a_2} presents the results
of employing the running-average technique with the window width
varied. While the correlation curve obtained in a standard manner
(top) does not contain any remarkable features and could be
attributed to virtually independent brightness variations, the
selection of short-term variations using windows as long as 39 min
(bottom panel) reveals a strong anticorrelation between the light
and the dark point with a nearly zero lag. Therefore, the
brightening events at one point of the pair nearly coincide in
time with darkening events at the other, while the sequence of
events is far from periodic, and no appreciable correlation can be
noted at lags other than zero. Thus, the updrafts of hot material
seem to be physically related to downdrafts of cool material,
although different couples of updraft and downdraft events appear
uncorrelated with one another on the time scales considered.
However, they are associated with the same region and should be
controlled by the same local system of convective circulation. A
correlation pattern of this type could arise if pairs of hot and
cool blobs are located in opposite parts of a streamline.

\begin{figure}[!h] 
\centering
   \includegraphics[bb=0 0 425 130pt,
   width=7.6cm,clip]{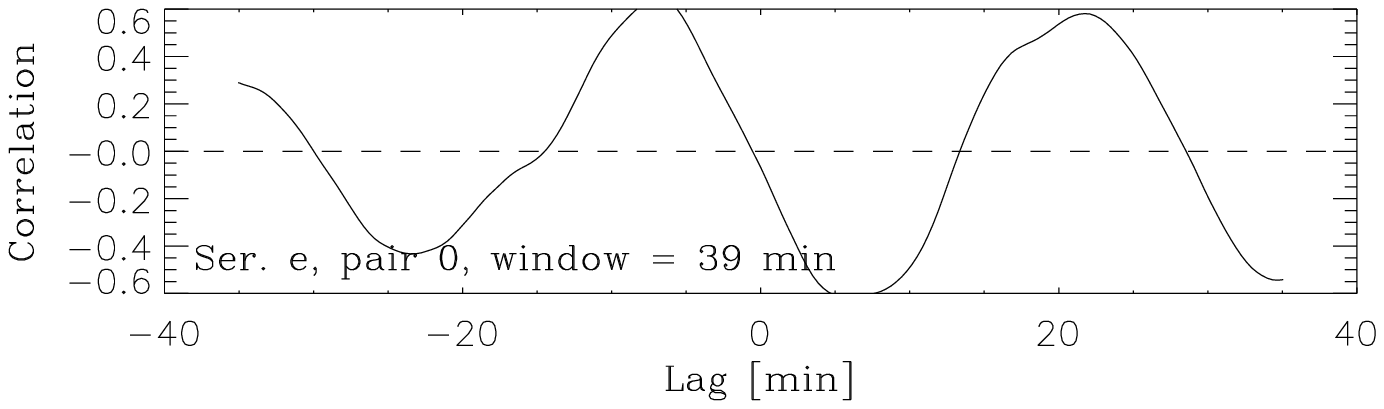}
   \includegraphics[bb=0 0 425 130pt,
   width=7.6cm,clip]{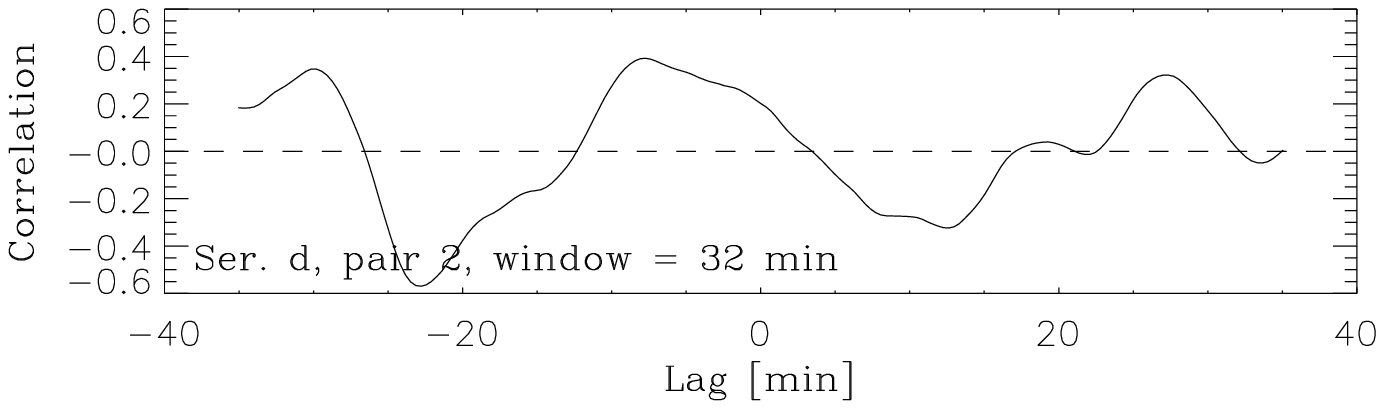}
\caption{%
Correlation curves obtained for two other pairs of points in the
ring system centred at point 0 (Figure \ref{5areas}) with a 39-min
(top) and a 32-min (bottom) running window.
}\label{ccs_32_d_2}\end{figure}

Even more interesting examples of correlation curves can be found
in Figure \ref{ccs_32_d_2}. For both the top and bottom panels,
the window widths (39 and 32 min, respectively) were chosen so as
to obtain maximum absolute values of the correlation at its
extrema. It can be seen that the two correlation functions are
qualitatively similar, although they refer to two different pairs
of points. The first one resembles a periodic function and
exhibits fairly large extremum values of the correlation, while
the second one is less regular. However, either has a negative
extremum between lag values of $-24$ and $-23$ min and a positive
one between $-8$ and $-7$ min. Moreover, the upper curve exhibits
a series of sign-alternating extrema, which follow at an interval
of 14--15 min. This pronounced tendency toward a periodic
variation in the correlation coefficient, which is especially
clear in the upper panel of Figure \ref{ccs_32_d_2}, can be
regarded as evidence for a nearly periodic reoccurrence of light
blobs (granules). If our interpretation is correct, we can use the
first curve to estimate the circulation period, which proves to be
29--30 min. In the second case, such a determination is less
certain, and the period may lie between 22 and 35 min.

Of course, the correlation curves are not always as regular as in
the above examples. It should be kept in mind, however, that the
choice of points was here almost arbitrary. In some cases, the
points chosen to form a pair might really belong to different
circulation systems (convection cells); in some others, they might
lie at widely diverging streamlines although in the same system;
finally, the circulation might locally be disturbed. However, the
very existence of pairs that demonstrate such patterns of
brightness correlation supports our suggestion that granules are
hot blobs carried by convective circulation, which can even
re-emerge on the photospheric surface.

\section{Rast's Comments on Paper I}
As we already mentioned, Rast \cite{rast} constructed a series of
random fields similar to granulation patterns in some respects
(mainly, in the mean values and standard deviations of certain
random parameters). All of them were additive superpositions of
$192^2$ randomly disposed two-dimensional Gaussians that randomly
varied in amplitude and radius around given mean values. These
values and the time scale of the evolution were chosen so as to
mimic closely the corresponding parameters of the solar
granulation. The sequence of such fields represented a continuous
time evolution with persistent emergence of ``new'' Gaussian peaks
and disappearance of ``old'' ones.

Averaging these fields over time revealed the following features.
First, isolated rectilinear chains of light blotches of
rectilinear dark lanes were discernible in some averages. Second,
the ``intensity''-variation curves for a local maximum and a
nearby local minimum of the averaged ``intensity'' sometimes
exhibited coincidence between a maximum of one curve and a minimum
of the other. Third, the rms contrast of the averaged fields
decreased with the averaging time in nearly the same manner as did
the contrast of the averaged granulation images.

Based on these three properties of the artificial fields, Rast
\cite{rast} claimed that the emergence of the quasi-regular
structures described in paper I and the apparent anticorrelation
between the intensity variations at the ``light'' and ``dark''
points are purely statistical rather than physical effects. Let us
discuss Rast's reasoning in the context of our findings.

First of all, the isolated linear features observed in Rast's
averaged artificial fields have very little in common with the
quasi-regular patterns revealed in the averaged images of the real
solar granulation. The quasi-regular trenching patterns are formed
by \emph{families} of ridges and trenches in the brightness
distributions --- parallel strips, concentric rings, and, in some
cases, radial ``spokes''. Isolated linear features can also be
found in the granulation images after time-averaging, but we make
little account of those mainly because they may indeed be of
statistical nature. In contrast, trenching patterns with signs of
spatial periodicity (completely absent in the artificial fields)
are of considerable interest from the standpoint of the
hydrodynamics of subphotospheric layers.

Next, the correlations presented in Figures \ref{ccs_varwin_a_2}
and \ref{ccs_32_d_2}, unlike the visually compared
brightness-variation curves, more definitely suggest that the
convective circulation could be a common physical factor
controlling the brightening events at one point and the darkening
events at the other --- in other words, the emergence of hot and
cool blobs. Moreover, the correlations nearly periodic as a
function of the time lag could naturally be attributed to the
recurrent emergence of light blobs (granules) carried by the
convective circulation. The characteristic circulation period of a
blob estimated as the period of variation of the correlation
coefficient is about two times as long as the characteristic
lifetime of granules (10--15 min), and this fact indirectly
supports our interpretation, according to which the granule should
be visible as it traverses the upper half of its closed trajectory
(streamline).

Finally, the important issue of the decrease in the rms contrast
of averaged images with the averaging time could not be resolved
based on the data presented in paper I. The spread in the original
contrast values for individual granulation images is very wide,
which makes it impossible to accurately compare the
contrast-decrease laws for the real granulation and for the
artificial random fields. As briefly noted by Brandt and Getling
\cite{bg} and will be shown more comprehensively in the
forthcoming paper by Brandt and Getling, the accuracy of such
comparisons can be substantially improved if the intensity of
images is properly renormalized, so as to make the rms contrast
and the mean intensity of each image equal to certain standard
values. We apply this normalization procedure to the real
granulation images of the La Palma series, to Rast's artificial
fields, and to two series of images obtained by numerically
simulating the solar granulation. We see that, for the averaged
granulation images, the curve of the contrast versus the averaging
time is flatter than for the remaining three series. Moreover, the
contrast of the averaged granulation images declines more slowly
than according to the $t^{-1/2}$ law typical of the averages of
random quantities ($t$ is the averaging time).

Thus, Rast's arguments do not appear to apply to the real
granulation patterns. Based on the above consideration, we can
dismiss Rast's criticism.

\section{Discussion and Conclusion}
We see that running-average movies constructed from both the La
Palma and SOHO MDI series of granulation images are very useful in
seeking long-lived, quasi-regular patterns. Families of straight
or circular rings and also spoke and web patterns no longer appear
to be unusual in time-averaged images of the solar granulation.
They can be distinguished most clearly if the averaging time is
about 1--2 h, so that their lifetimes are at least of this order
of magnitude. Thermal convection of a Rayleigh--B\'enard type is
rich in a variety of flow patterns that it can form (see, e.g.,
Getling (1998) for a survey), and this physical mechanism could
underlie the pattern-forming processes in the solar
subphotospheric layers. It is worth remembering in this context
that families of concentric rings distinguishable in some averaged
photospheric images resemble the so-called target patterns
observed in experiments on Rayleigh--B\'enard thermal convection
(see Assenheimer and Steinberg (1994) and a reproduction of their
experimental photograph in paper I).

Applying our azimuthal-averaging algorithm to time-averaged images
highlights, for some centres, systems of concentric annuli with
substantially enhanced mean intensities. An algorithm of
aerospace-image processing, tentatively employed to analyse
time-averaged granulation images, can also be useful for the
detection of trenching patterns in the brightness distributions.

Our analysis of correlations between brightness variations at the
points corresponding to a local intensity maximum and a nearby
intensity minimum in an averaged image suggests that blobs of hot
material (which appear as granules) can repeatedly emerge on the
photospheric surface, and their lifetime may be considerably
longer than the lifetime of an individual observed granule.
Another manifestation of this process could be the repeated
expansion and fragmentation of granules associated with centres
where the horizontal-velocity field has a strong positive
divergence (M\"uller {\it et al.}, 2001).

In view of the properties of granulation patterns described here
and with due account for the contrast-variation laws reported by
Brandt and Getling \cite{bg} and planned to be analysed in more
detail in the companion paper, we can dismiss the critical
comments to paper I expressed by Rast \cite{rast}.

As we noted in paper I, signs of the prolonged persistence of
granulation patterns were observed previously. Roudier {\it et
al.} \cite{roudier} detected long-lived singularities (dark
features --- ``intergranular holes'') in the network of
supergranular lanes. They were continuously observed for more than
45 min, and their diameters varied from 0.24$\arcsec$ (180 km) to
0.45$\arcsec$ (330 km). Later, Hoekzema {\it et al.} \cite{hbr}
and Hoekzema and Brandt \cite{hb} also studied similar features,
which were observable for 2.5 h in some cases.

An interesting parallel to our observation of concentric-ring
patterns was reported by Berrilli {\it et al.} \cite{bdm-str}, who
analysed pair correlations in the supergranular and granular
fields. They defined the pair-correlation function $g_2(r)$ so
that the probability of finding a target supergranule or granule
(identified by its barycentre) within an annulus of radius $r$ and
width $\mathrm dr$ is $2\pi r g_2(r)\mathrm dr$, where $r$ is
measured from the barycentre of a chosen reference supergranule or
granule. Then the positions and heights of the peaks of such a
function should reflect the topological order in the system. The
computations of $g_2(r)$ based on observed supergranular and
granular fields yielded qualitatively similar oscillating
functions whose local amplitudes (measured from the mean $g_2(r)$
values) decrease with $r$ in each case from a maximum at some
small $r$. This means that the patterns of supergranular and
granular barycentres are most ordered at small $r$ and become less
ordered at larger $r$. Such a behaviour of $g_2(r)$ could
naturally be expected for supergranules, which form closely-packed
patterns. In contrast, the presence of a similar (although not so
perfect) order in the granulation field, especially pronounced in
time-averaged granulation images, is much less trivial. (We note,
however, that the authors themselves emphasize differences rather
than similarities between the supergranular and granular fields in
their topological orders.)

To all appearances, the interpretation of the quasi-regular
patterns observed in the averaged granulation images should be
based on analyses of the stability properties of cellular
convection on meso- and supergranular scales. Variations in the
local intensity of the convective circulation can affect the
thermal structure of subphotospheric layers and, eventually, the
fine structure of the velocity field in these layers. We plan to
discuss these issues elsewhere.

\begin{acknowledgements}
I am grateful to P.N. Brandt for his cooperation and hospitality
during my visits to the Kiepenheuer-Institut f\"ur Sonnenphysik
(Freiburg), for numerous, extensive discussions of the subject,
and for his comments on this paper; to R.A. Shine for making
available the SOHO MDI data; to A.A.~Buchnev and V.P. Pyatkin
(Institute of Computational Mathematics and Mathematical
Geophysics, Novosibirsk) for considering possible adaptations of
the aerospace-image-processing software for analyses of solar
granulation patterns and to A.A. Buchnev for the tentative
processing of some images; to D. Del Moro for the discussion of
the correlation properties of the supergranular field revealed by
Berrilli {\it et al.} \cite{bdm-str}; and to a referee and to L.M.
Alekseeva for useful remarks. This work was supported by the
\emph{Deut\-sche For\-schungs\-ge\-mein\-schaft}\ (project 436 RUS
17/56/03) and by the Russian Foundation for Basic Research
(project 04-02-16580-a).
\end{acknowledgements}

\end{article}
\end{document}